\documentclass[12pt]{article}
\usepackage{sc3conf}
\usepackage{amsfonts}
\usepackage{epsfig}

\newcommand{\bee}{\begin{eqnarray}}
\newcommand{\eend}{\end{eqnarray}}
\newcommand{\fmin}{f_{-\lambda,k}(r)}
\newcommand{\f}{f_{\lambda,k}(r)}

\newcommand{\fk}{f_{\lambda,-k}(r)}
\newcommand{\pmin}{\phi_{-\lambda,k}(r)}

\newcommand{\p}{\phi_{\lambda,k}(r)}
\newcommand{\pk}{\phi_{\lambda,-k}(r)}
\newcommand{\Sch}{{Schr$\ddot{\rm o}$dinger equation }}
\newcommand{\la}{\lambda}

\newcommand{\il}{{\rm {Im}}\la}
\newcommand{\rmd}{{\rm d}}
\newcommand{\rme}{{\rm e}}
\newcommand{\rmi}{{\rm i}}
\begin{document}

\title{Black-hole approach to the singular problem of quantum mechanics.II
}

\authors{A.E.~Shabad}

\addresses{P.N.Lebedev Physics Institute, Russian Academy of Sciences,
Leninsky prospect 53, Moscow, Russia.}

\maketitle

\begin{abstract}
A new approach is proposed for the quantum mechanical problem
 of the falling of a particle to a singularly attracting center,
 basing on a black-hole concept of the latter.

 The singularity $\sim r^{-2}$ in the potential of the radial \Sch
 is considered as an emitting/absorbing center. The two solutions
 oscillating in the origin are treated as asymptotically free
 particles, which implies that the singular point $r=0$ in the \Sch is
 treated on the same physical ground as the singular point $r=\infty$.
 To make this interpretation possible, it is needed that the
norm squared of the wave function $\int_r|\psi(r)|^2\rmd\mu(r)$
should $diverge$ when $r\rightarrow 0$, in other words, the
measure used in definition of scalar products should be singular
in the origin. Such measure comes into play if the \Sch is written
in the form of the generalized (Kamke) eigenvalue problem for
either of two - chosen differently depending on the sign of the
energy $E$ - operators, other than Hamiltonian. The Hilbert spaces
where these two operators act are used to classify physical
states, which are: i) states of "confinement"- continuum of
solutions localized near the origin, $E<0$ - and ii) the states
corresponding to the inelastic process of reflection/transmission,
$i.$ $e.$ to transitions in-between states localized near the
origin and in the infinitely remote region, $E>0$. The
corresponding unitary $2\times2$ $S$-matrix is written in terms of
the Jost functions. The complete orthonormal sets of
eigen-solutions of the two operators are found using "quantization
in a box" ($r_L,~r_U)$, followed by the transition to the limit
$r_L\rightarrow 0$, $r_U\rightarrow\infty$. The corresponding
expansions of the unity are written.
\newpage
\end{abstract}
\section{Introduction}
We study the radial \Sch \begin{eqnarray}\label{4}
H\psi(r)=k^2\psi(r)
\end{eqnarray}
\begin{eqnarray}\label{5}
H=-\frac{\rmd^2}{\rmd r^2}+\frac{\lambda^2-\frac 1{4}}{r^2}+V(r),
\qquad 0\leq r<\infty,
\end{eqnarray}
where $k^2=E$ is the energy, and $\lambda$, when taken imaginary,
is a coupling constant of singular attraction. This parameter may
be also thought of as connected with "complex angular momentum"
$l=\lambda-\frac1{2}$. The potential $V$ is assumed real and
well-behaved, so
 as no extra troubles be introduced:
\begin{eqnarray}\label{poten}
\int_c^\infty|V(r)|\rmd r<\infty,  \quad \int_0^{c'}r|V(r)|\rmd
r<\infty, \quad c,c'>0.
\end{eqnarray}
The case of $V(r)=0$ is explicitly solvable in terms of
cylindrical functions.

The problem under consideration is usually addressed with the help
of self-adjoint extension of the Hamiltonian (\ref{4}). This was
first done by K.Meetz \cite{meetz} (see  references to later works
in the most recent publication \cite{gavrilov}), who grounded in
this way the conjecture of K.M.~Case \cite{case}, (see also
\cite{morse}). The spectrum of the Hamiltonian,~ accurately
defined via the extension procedure within the von Neuman theory
(see e.g.\cite{naimark} and the comprehensive physical survey in
\cite{bonneau} ), is discrete and unlimited from below.

This circumstance is often recognized as physically unsatisfactory
and became the motivation for issuing a different approach
\cite{shabad}, which respects the principle of correspondence with
classical mechanics, where a particle, placed in the field of the
singular center,
 performs a spiral motion diverging from the center or converging towards
it and making an
 $infinite$ number of revolutions around it.  A quantization, taking
 into account the
 correspondence principle, should describe a center, which emits and
 absorbs particles that are free  near the origin,
 since in the classics the motion of a particle is sort of unbounded in
  this vicinity,
 similarly to their motion in the region remote to infinity. In short, the
 most characteristic feature of the approach of Ref.\cite{shabad},
 which we are continuing in the present publication,
  is that, in it,
  mathematical conditions are provided to make it possible to treat
  the singular point $r=0$ of the differential equation (\ref{4}),
(\ref{5}) on
  the same physical
 footing, as the singular point $r=\infty$ is usually treated.

 To be more precise,  instead
of $H$, eq.(\ref{5}), we use two  differential operations $H^{IV}$
and $H^{III}$
\begin{equation}\label{HIV}
H^{IV}=-\frac{\rmd^2}{\rmd r^2}-\frac{1}{4 r^2}+V(r),\quad
\end{equation} and \begin{equation}\label{HIII}
H^{III}=-\frac{\rmd^2}{\rmd r^2}-\frac{1}{4 r^2}+V(r)-k^2,
\end{equation}
(one appropriate, if $E>0$, and the other, if $E<0$) for
classifying
 physical states as vectors in the Hilbert spaces of the correspondingly
 defined self-adjoint operators.
 These operators are associated with the same equation
 (\ref{4}),
 (\ref{5}), but do not have the above unwanted properties of the Hamiltonian.
 The \Sch (\ref{4}), (\ref{5}) becomes a generalized eigenvalue problem
of the type, studied by Kamke \cite{Kamke}, for either of these
operators:\bee\label{x}
H^{IV}\psi_{\la,R}(r)=(\il)^2\left(\frac{1}{R^2}+\frac
1{r^2}\right)\psi_{\la,R} (r),\nonumber\\
R=\frac{-\il}{k},\quad k^2>0,\quad\la^2<0\quad{\rm (consider} R>0 {\rm
~for~ definiteness)} \eend and
\begin{equation}\label{III}
H^{III}\psi_{\la,k} (r)=\frac{(\il)^2}{r^2}~\psi_{\la,k} (r),\quad
k^2<0,\quad\la^2<0.\end{equation} In eq.(\ref{III}) the eigenvalue
is $(\il)^2$, with the energy $k^2$ kept fixed as a negative
parameter. In eq.(\ref{x}) the eigenvalue is $(\il)^2$, with the
ratio $R=-\il/k$ kept as a real parameter. Alternatively, writing
the factor in front of $\psi$ in the r.-h. side of eq.(\ref{x}) in
the form $k^2(1+R^2/r^2)$, one may consider  the energy $k^2$ as an
eigenvalue, with $R$ being a fixed parameter.
 These problems introduce new
measures\bee\label{measureIV} \rmd\mu^{IV} (r)=
\left(\frac{1}{R^2}+\frac{1}{r^2}\right)\rmd r,\eend
\bee\label{muIII} \rmd\mu^{III}(r)=\frac 1{r^2}~\rmd r, \eend
which are to be used later in defining scalar products in the
corresponding Hilbert spaces of eigenvectors of $H^{IV}$ and
$H^{III}$. The measures (\ref{measureIV}) and (\ref{muIII}) are
both singular in the origin $r=0$. This fact is of crucial
importance for providing the possibility of treating the
solutions, oscillating in the origin, as corresponding to free
particles. We comment on this point a little later.

 For defining $H^{IV}$ and $H^{III}$ as self-adjoint operators,
 and studying their spectra
we use the physically  straightforward procedure
 of quantization in a box.  When  doing so we introduce the lower
  boundary
$r_L$ of the box and let it later tend to the point of singularity
of the differential equation $r_L\rightarrow 0$, exactly in the
same manner, as we introduce the upper boundary $r_U$ and let it
tend to the other singularity point $r_U\rightarrow\infty$, in
accord with the customary procedure. We impose zero boundary
conditions in the point $r=r_L$ for problem (\ref{III}), and
(anti)periodic boundary conditions at the ends of the box
$r=r_L,r_U$ for problem (\ref{x}). The spectra, discrete as long
as $r_L$ and $r_U$ are finite, turn into continua in the limit
$r_L=0$, $r_U=\infty$.

Certainly, for a self-adjoint definition of the operators $H^{IV}$
and $H^{III}$, the von Neuman technique
 might be applied, as well. As long as the finite interval
 $(r_L,r_U)$ is concerned, this makes no difference with the
 quantization in a box.
  The special power
 of von Neuman technique lies in its ability to handle directly the
 intervals
  with singular ends. We insist, however, that the corresponding
 results may make physical sense only to the extent, to which these
 reproduce the limit $r_L\rightarrow 0$, $r_U\rightarrow\infty$,
 since, in the physical reality,
 there are not infinitely large boxes, as well as there are not
 point-like sources of force: there may only be very large boxes and
 very small-sized sources (as compared to other parameters with the
 dimensionality of length involved in the problem).

Transformations are known \cite{Kamke},~\cite{naimark}( see eqs.
(\ref{21}),(\ref{22}) and eqs. (\ref{35}),(\ref{36}) below) that
reduce the Kamke eigenvalue problems (\ref{III}), (\ref{x}) to the
standard Liouville forms ( see eq. (\ref{24}) and eq. (\ref{37})
below), to which the  theory of self-adjoint differential
operators is directly applicable.

Both of the transformations in the asymptotic region $r\rightarrow
0$ reduce to the change
  of the variable ( $r_0$ is a free dimensional parameter)
 \begin{equation}\label{21}
r_*=r_0\ln\frac r{r_0},
\end{equation}
 accompanied with the transformation of the wave function
\begin{equation}\label{22}
\left(\frac{r_0}{r}\right)^\frac1{2}\psi(r)=\tilde{\psi}(r_*).
\end{equation}
The origin $r=0$ is mapped onto $r_*=-\infty$. The singularity in
the point $r=0$ gives rise to solutions with the oscillating
 asymptotic behavior at $r\rightarrow 0$, $r_*\rightarrow -\infty$
 \begin{eqnarray}\label{1.2}
\psi(r)\asymp r^{\pm\rmi\rm{Im}\lambda+\frac 1{2}} \eend
\bee\label{1.3} \tilde{\psi}(r_*)\asymp\exp{(\pm\rmi r_*\frac{\rm
{Im}\lambda}{r_0})}r_0^{\rmi{\rm Im}\lambda}.
\end{eqnarray}
The latter equation is a free wave. This observation alone is not
yet enough to make (\ref{1.3}) correspond to a free particle. To
do this consider the behavior of the norm, associated with the
transformed~\Sch (see Sections {\bf 2} and {\bf 3} below), at the
lower integration limit
 \bee\label{1.4}
 \int_{-L}|\tilde{\psi}(r_*)|^2\rmd r_*=r_0^2\int_{r_0\exp (-L/r_0)}|\psi (r)|^2
 \frac{\rmd r}{r^2}.
 \eend
 Eq.~(\ref{1.4}) diverges linearly with the box size in the $r_*$-space
 $L=r_0\ln (r_0/r_L)\rightarrow\infty$.
 This is just what is needed to argue that the particle spends most part of its
 life in the form of a free particle in the asymptotic region
 $r\rightarrow 0$, the singularity in the integration
 measure $\rmd r/r^2$ in (\ref{1.4}) providing a sufficiently ample volume for doing this.
 This statement establishes the more precise meaning of the phrase~\cite{QM}
 "falling of a particle down to the center". Unlike~\cite{QM} we do not attribute
 this phenomenon to the fact that there is a point $E=-\infty$ in the energy spectrum.
 On the contrary,
the states of $H^{III}$, asymptotically free near the origin, make a continuum
 (the situation is the same
 as for the usual continuum of states free at $r=\infty$ for real $k$).

After reduced to the standard Liouville form (see eq. (\ref{24})
below), problem (\ref{III}) describes particles, issued at
negative infinity and totally reflected by impenetrable potential,
resulting from the above transformation, back to negative
infinity. This is a complete analogue of the elastic scattering of
particles, belonging to the continuous spectrum of $H$, but now in
the "inner" world near the singularity. The scattering is
characterized by one scattering phase. The probability flux to or
from the singularity is zero. We refer to this situation that
occurs in the domain $k^2<0,$ $\la^2<0$, called sector III in
\cite{shabad}, as $inner ~elastic~ scattering ~in
~continuum~of~asymptotically$-$free~confined~states$, since the
eigenfunctions of $H^{III}$ are concentrated near the origin
$r=0$, belong to a continuum  and oscillate like a free
exponential (with diverging norm) when approaching the point
$r=0$.

In contrast to sector III, the eigenvalue problem (\ref{x}),
appropriate in the domain $k^2>0,$ $\la^2<0$, called sector IV in
\cite{shabad}, becomes, after the corresponding transformation
reduces it to the standard Liouville form (see eq. (\ref{37})
below), a barrier penetration problem on the whole axis
$(-\infty,\infty)$. Particles, incoming from positive (negative)
infinity are partially reflected back by the barrier potential,
 and partially penetrate through this potential to
outgo to negative (positive) infinity. In accord with the
(anti)periodic boundary conditions, imposed in problem (\ref{x}),
the total probability flux is, generally, nonvanishing. This
confinement/deconfinement process - we call it this way, because
particle outgoing to or incoming from the negative infinity
represent transitions to or from the continuum of asymptotically
free, $\delta$-function normalized states, confined by the center
- is described by a $2\times 2$ unitary scattering matrix of an
inelastic interchannel process, determined by two scattering
phases and one inelasticity angle.

In Sec.~\ref{conf.} the spectral problem (\ref{III}) is studied in
sector III,  the complete set of  orthonormal eigenfunctions is
found in a finite box and in the continuum limit, as well as the
scattering phase for the elastic scattering of particles, emitted
by the center - back to the center. In Sec.~\ref{Two-chan} we
fulfil the same program for the spectral problem (\ref{x}) in
sector IV. The complete orthonormal set of eigenfunctions is
found, which behave as standing wave both near the center and near
infinity, the corresponding expansion of unity is written. The
$2\times2$  scattering matrix  elements are expressed in terms of
the Jost functions.

\section{Continuum of confined states}\label{conf.}
In the  domain of parameters $\la^2<0$, $k^2<0$, called sector III
in~\cite{shabad}, only one fundamental solution is appropriate,
the boundary conditions are to be imposed at one end-point and
belong to Sturm-Liouville type, the probability flux is zero.

Define a new differential operatation  $H^{III}$ according to eq.
(\ref{HIII}) so that the \Sch  (\ref{4})  take the form
(\ref{III}). We consider this equation in the half-box
\begin{equation}\label{half}
r_L\leq r<\infty.
\end{equation}
The lower limit of the half-box $r_L$ is meant to tend to zero
afterwards. The differential equation (\ref{III}), defined on the
interval  (\ref{half})  make the $general$ $eigenvalue$ $problem$
of Kamke~\cite{Kamke}, for which it is peculiar that the
eigenfunction in the r.-h. side is taken with a variable-depending
factor, here $\lambda^2/r^2$, not just the constant eigenvalue
$\lambda^2$. If boundary conditions are chosen in such a way that
$H^{III}$ is symmetric (Hermitian), it is also self-adjoint and
the set of eigenfunctions of the problem (\ref{III})  is complete
in a Hilbert space. To see this, it is sufficient to perform the
transformation, which reduces the problem (\ref{III}) to the
normal Liouville form. However, the orthonormality of the
eigenfunctions with nonflat measure can be derived already before 
we make this
transformation.

Impose the "zero boundary conditions" at the 
both ends of the
half-box
\begin{equation}\label{bound}
\psi (\infty)=0,   \qquad            \psi (r_L)=0.
\end{equation}
The first one means in fact that the domain $\mathcal{D}_{III}$
where $H^{III}$ acts consists of functions $\psi(r)$ that decrease
fast enough when $r\rightarrow\infty$. This implies that only one
- out of the two -  fundamental solutions is appropriate,  the
deficiency index for the problem (\ref{III}) with one regular,
$r=r_L$, and one singular $r=\infty$ end being $m_{r_L,\infty}=1$
. Physically, the zero boundary conditions (\ref{bound}) guarantee
that the overall probability flux carried by functions
$\psi\in\mathcal{D}_{III}$ \bee\label{27a}
 P_\psi=\rmi\left(\psi(r)\frac{\rmd\psi^*(r)}{\rmd r}-
      \psi^*(r)\frac{\rmd\psi (r)}{\rmd r}\right)
\eend
 to/from the center $r = r_L$ be zero, as well as the probability
 flux to/from the infinity $r = \infty$. This holds true also when
$r_L \rightarrow 0$. We discuss later, whether the second
condition in (\ref{bound}) is the most general choice or not.

The operator  (\ref{HIII}) is symmetric
(Hermitian)\begin{equation}\label{self} (H^{III})^*=(H^{III})^T,
\end{equation}
provided that its matrix element 
is defined as
\begin{equation}\label{matrix}
H_{i j}^{III} = \int_{r_L}^\infty \psi_i^*(r) H^{III} \psi_j
(r)\rmd r,
\end{equation}
where   $\psi_{i,j}(r)$ are any two square-integrable functions,
sufficiently smooth in the interval  (\ref{half}), subject to
conditions (\ref{bound}). The asterisk in   (\ref{matrix})
designates complex conjugation, and $T$ indicates transposition.
(Remind, that $V(r)$, $ k^2$ are both real.)

As long as the lower box end $r_L$ is finite, the Kamke eigenvalue
problem (\ref{III}), (\ref{bound}) has a discrete spectrum. When
$r_L\ll r_0$, where $r_0$ is a dimensional parameter, the discrete
eigenvalues of $H^{III}$ behave like~\cite{shabad} $\la_n = \rmi
\pi n(\ln (r_0/r_L))^{-1}$ and do not depend on $k$. (We come back
to this point below in this section). In the limit $r_L=0$ they
condense~\cite{shabad}, as is usually the case with quantization
in a box, and we are left with a continuum of states, which we
call confined, since the  functions from $\mathcal{D}_{III}$ are
concentrated in a finite domain.

Owing to the Hermiticity property (\ref{self}), any two solutions
$\psi_{\la_{1,2},k}$ of the Kamke eigenvalue problem (\ref{III}),
(\ref{bound}) obey the relation
\begin{equation}\label{orthog}
(\la_1^2-\la_2^2)\int_{r_L}^\infty\psi^*_{\la_1,k}(r)\psi_{\la_
2,k}(r)\frac{\rmd r}{r^2} = 0,
\end{equation}
which implies that these be orthogonal with the measure $\rmd
r/r^2$, provided the (real) eigenvalues $\la^2$ are different,
$\la_1^2\neq\la_2^2$, while the energy $k^2$ is the same. The
equality of  the $k$'s in the two functions $\psi^*_{\la_1,k}(r)$
ans $ \psi_{\la_ 2,k}(r)$ in (\ref{orthog}) is dictated by the
fact that $H^{III}$, eq. (\ref{HIII}), contains $k$.  The
derivation of (\ref{orthog}) is standard: one should left-multiply
the equation\begin{equation}\label{la1} H^{III}\psi_{\lambda_1,~k
}^*(r)=-\frac{\lambda_1^2}{r^2}\psi_{\lambda_1,~k }^*(r)
\end{equation}
by $\psi_{\lambda_2, k }(r)$, and the equation
\begin{equation}\label{la2}
H^{III}\psi_{\lambda_2, k
}(r)=-\frac{\lambda_2^2}{r^2}\psi_{\lambda_2, k }(r).
\end{equation}
by $\psi_{\lambda_1, k }^*(r)$. The difference of these products,
when integrated over $\rmd r$, vanishes due to (\ref{self}) to
give (\ref{orthog}).
 The ortho-normality relations,
which follow from (\ref{orthog}) in the continuum limit $r_L=0$
are ~\cite{shabad} \bee\label{48}
\kern-50pt\frac{2(\il)^2}{|f(\la,-k)|^2}\int_0^\infty
f_{\la,-k}(r)f^*_{\la',-k}(r)\frac{\rmd r}{r^2} =\pi\delta({\rm
Im}\la-{\rm Im}\la'). \eend Here  $\fk$ designates the exact
solution to the \Sch (\ref{4}), valid in the whole domain $r\in
(0,\infty)$, which decreases for $r\rightarrow\infty$ as
exp$(-r{\rm Im}k)$ (consider Im$k>0$ for definiteness), and
$f(\la,k)$ is the Jost function, defined as the Wronsky
determinant~\cite{regge}
\begin{eqnarray}\label{11}
f(\lambda,k)=\f\frac{\rmd\p}{\rmd r}-\frac{\rmd\f}{\rmd r}\p.
\end{eqnarray}
between the solution $\f$ and another solution, called $\p$, which
oscillates like $r^{\rmi{\rm Im\la}+1/2}$ near $r = 0$. For the
case $V=0$ eq.(\ref{48}) becomes
 an orthogonality relation for McDonald functions $K_{\rmi{\rm Im}\la}
 (r)$ with (different or coinciding) imaginary indices
$K_{\rmi{\rm Im}\la}(r)$ (see~\cite{shabad}).

Now it is time to perform the advertised transformation. This is
the transformation
  of the coordinate (\ref{21})  and of the wave
function~(\ref{22}) made in the whole domain (\ref{half}) (not
only near $r=0$, as it was discussed in Introduction).
 After this transformation, the \Sch
(\ref{4}) or (\ref{III}) aquires the standard Liouville form
  \begin{equation}\label{24}
\kern-40pt\left(-\frac{\rmd^2}{\rmd r_*^2}+U_{\rm
{con}}(r_*)\right)\tilde{\psi}(r_*) =\frac{(\rm
Im\la)^2}{r_0^2}\tilde{\psi}(r_*)
\end{equation}
with
\begin{equation}\label{25}
U_{\rm{con}}(r_*)=-\exp\left(\frac{2r_*}{r_0}\right)
\left(k^2-V(r_0\exp\frac{r_*}{r_0})\right).
\end{equation}
The transformation  (\ref{21}) of the coordinate  maps the
half-box (\ref{half})  to the half-box
\begin{equation}\label{half2}
-L\leq r_*<\infty,\quad r_L=\rme^{-\frac L{r_0}},
\end{equation}
where $L = r_0\ln (r_0/r_L)$. The left wall $r_*=-L$ of the
half-box (\ref{half2})
 in the $r_*$-space tends to negative infinity as  the core radius tends to zero,
$(r_L/r_0)\rightarrow 0$. The boundary conditions (\ref{bound})
now become
\begin{equation}\label{bound2}
\tilde{\psi}(\infty)=0,   \qquad      \tilde{\psi}(-L)=0.
\end{equation}
The transformation of the wave function (\ref{22}) is intended to
meet the requirement
 that there should be
no linear-derivative term in (\ref{24}). Equation (\ref{24}) with
the boundary conditions (\ref{bound2}) has the form of a usual
eigenvalue problem, with the potential $U_{con}(r_*)$ containing
$k^2<0$ as a parameter. It
 proposes the customary measure $\rmd r_*$ to be used in defining the
norm. The following relation between the norms in the $r$- and
$r_*$-spaces \bee\label{norms}
 \int_{-L}^\infty|\tilde{\psi}(r_*)|^2\rmd r_*=
r_0^2\int_{r_L}^\infty|\psi (r)|^2
 \frac{\rmd r}{r^2}
 \eend
takes place.

\begin{figure}[htb]
  \begin{center}
   \includegraphics[bb = 0 0 405 210,
    scale=1]{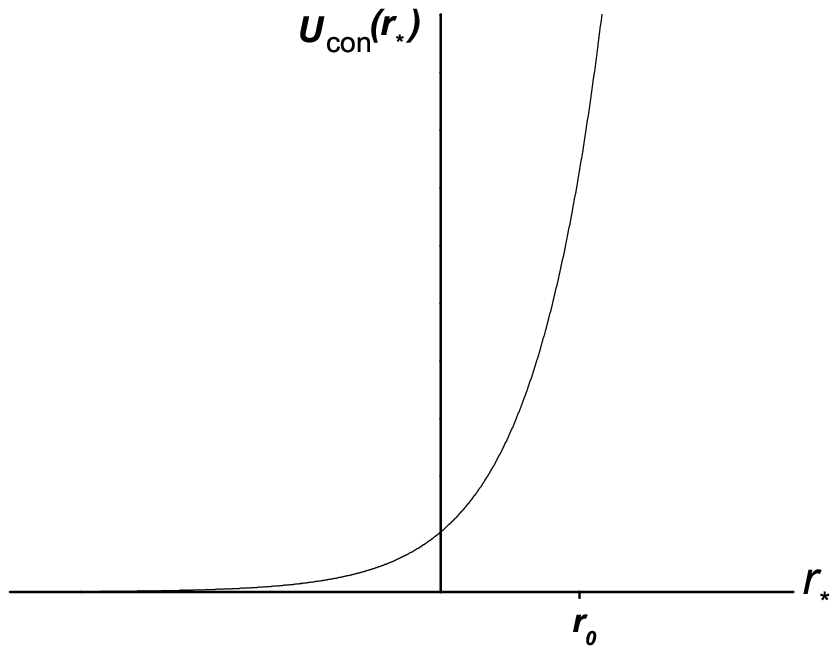}
    \caption{The effective potential in the confining sector for the
    case of $V=0$ (see eq.~(\ref{25})). It prevents particles free near the singularity
     $r_*=-\infty~ (r=0)$ from escaping to the outer world ($r_*\gg r_0$).
      }\label{fig:1}
  \end{center}
\end{figure}

The effective potential $U_{\rm{con}}(r_*)$ is plotted in
Fig.~\ref{fig:1}  for the  case of $V=0$.

 The inclusion of $V\neq 0$ cannot change the asymptotic forms
$U(-\infty)=0$, $U(\infty)=\infty$ due to the condition
(\ref{poten}). Solutions of eq.~(\ref{24}) are free waves
(\ref{1.3}) near the negative infinity and are totally reflected
by the effective potential (\ref{25}) to the left side. This
strictly forbids their penetration into the outer world
$r_*\rightarrow\infty$. We face the process of elastic scattering
of particles, incoming from the negative infinity of the
$r_*$-axis ($i.e.$, emitted by the center) back to the negative
infinity ( to be absorbed by the center). The free parameter of
dimension of length $r_0$ plays the role of the size of the
system. As mentioned before, the total probability flux is zero.
This means that the center absorbs all what it emits. Note, that
the flux (\ref{27a}) is invariant under the transformation
(\ref{21}),~(\ref{22}), $i.e.$, it does
  not change if one replaces $\psi (r)$ by $\tilde{\psi}(r)$ and
$r$ by $r_*$.

This elastic scattering process in the inner world may be
described exactly in the same terms as the usual one. The solution
$\fk$, defined above, is a linear combination
\begin{eqnarray}\label{b}
\fk=C\p+D\pmin
\end{eqnarray}
of two solutions,  $\phi_{\pm\la,k}(r)$, that oscillate
like~(\ref{1.2}) near $r=0$. The coefficients $C$ and $D$ here are
expressed ~\cite{regge} in terms of the Jost function (\ref{11}).
Define the dimensionless solution $\overline{\phi}_{\pm\la,k }(r)$
\bee\label{phibar}\overline{\phi}_{\pm\la,k}(r)=|k|^{\pm\rmi\il+1/2}
\phi_{\pm\la,k}(r)\asymp
~_{r\rightarrow\infty}~|rk|^{\pm\rmi\il+1/2}\eend and its Wronsky
determinant with $f_{\pm\la,k}(r)$ \bee\label{fbar} \overline{f}
(\pm\la,k)=|k|^{\pm\rmi\il+1/2}f(\pm\la,k).\eend The
dimensionality of $\overline{f}$ is $[length]^{-1}$.
 Then (\ref{b}) is represented as \bee
\fk=\frac{\overline{f}(\la,-k)}{2\la |k|}~ [\exp
(\rmi\delta^{III})~\overline{\phi}_{\la,k}(r)+\overline{\phi}_{-\la,k}(r)],
\eend where \bee\label{33}
\delta^{III}(\la,k)=\pi-2\arg\overline{f}(\la,-k). \eend

 The boundary
conditions (\ref{bound}) or (\ref{bound2}) are satisfied provided
that the  spectral equation \bee\label{spect} \la_n=\rmi
n\pi\frac{r_0}{L}+\frac{r_0}{L}\left( \la_n\ln |kr_0|-\rmi
\arg\overline{f}(\la_n,-k)\right),\quad n=0,\pm 1,\pm 2... \eend
is solved with $n$ being an arbitrary integer. When
$L\rightarrow\infty$ the second term in (\ref{spect}) should be
neglected and the spectrum becomes $\la_n=\rmi \pi n r_0/L$, as
stated above.

The normalized solution of the eigenvalue problem (\ref{III}),
(\ref{bound})\bee\label{normsolution}
\psi_{\la,k}(r)=\frac{|\il|\sqrt
{2}}{\sqrt{\pi}|f(\la,-k)|}f_{\la,-k}(r)\eend has the
form\bee\label{normsolution2}
\psi_{\la,k}(r)=\frac{\rmi}{\sqrt{2\pi}} \left(\rme^{-\rmi\arg
f(\la,-k)}\phi_{\la,k}(r)-\rme^{\rmi\arg
f(\la,-k)}\phi_{-\la,k}(r)\right)=\nonumber\\
=\frac{\rmi}{\sqrt{2\pi|k|}} \left(\rme^{-\rmi\arg
\overline{f}(\la,-k)}\overline{\phi}_{\la,k}(r)-\rme^{\rmi\arg
\overline{f}(\la,-k)}\overline{\phi}_{-\la,k}(r)\right).\eend Near
the singularity point $r=0$ the eigenfunctions behave
as\bee\label{near=0} \left.\psi_{\la,k}(r)\right|_{r\rightarrow
0}\asymp\frac{\sqrt{2r}}{\sqrt{\pi}}\sin
\left(\arg\overline{f}(\la,-k)-\il\ln(r|k|)\right)\eend
Correspondingly, the eigenfunctions
$\widetilde{\psi}_{\la,k}(r_*)$ of the problem (\ref{24}) -
(\ref{bound2}), which are the functions (\ref{normsolution}),
transformed according to (\ref{21}), (\ref{22}), behave near the
singularity point $r_*=-\infty$ as\bee\label{near-ifty}
\left.\widetilde{\psi}_{\la,k}(r_*)\right|_{r_*\rightarrow-\infty}
\asymp\frac{\sqrt{2r_0}}{\sqrt{\pi}}\sin
\left(\arg\overline{f}(\la,-k)-\il\ln(r_0|k|)-\il\frac{r_*}{r_0}
\right).\eend For $V=0$, the dimensionless  phase
$\arg\overline{f}(\la,-k)$ cannot depend on $k$, since  there is
no other dimensional parameter in the \Sch (\ref{4}) or
(\ref{III}) in this special case. From the known exact solution of
the \Sch one finds: \bee\label{freephaseIII}
\arg\overline{f}_0(\la,-k)=-\arg\Gamma(1-\rmi\il) \eend

The form (\ref{near-ifty}) proposes the definition of the
scattering matrix (just a unit-length complex number) describing
the internal elastic scattering:\bee\label{SIII}
S^{III}=\rme^{2\rmi(\arg\overline{f}(\la,-k))}(r_0|k|)^{-2\rmi\il}\eend

The orthonormality relations (\ref{48}), written for the
eigenfunctions $\widetilde{\psi}_{\la,k}(r_*)$ of the problem
(\ref{24}) - (\ref{bound2}) or for (\ref{normsolution}), take the
form\bee\label{ort} \frac
1{r_0^2}\int_{-\infty}^\infty\widetilde{\psi}_{\la,k}(r_*)
\widetilde{\psi}^*_{\la',k}(r_*)\rmd r_*=
\int_{0}^\infty{\psi}_{\la,k}(r) {\psi}^*_{\la',k}(r)\frac{\rmd
r}{r^2}=\delta\left({\rm Im}\la -{\rm Im}\la'\right). \eend Note,
that as long as the parameter Im$\la$ can be viewed upon as a
strength of the singular attraction, the ortho-normality relation
(\ref{ort}) expresses spectral properties with respect to a
"coupling constant".

The eigenfunctions (\ref{normsolution}) or (\ref{normsolution2})
that belong to the continuum make a complete system, unless the
eigenvalue problem (\ref{24}) - (\ref{bound}) has extra discrete
solutions ( this depends upon the potential $V(r)$) for negative
values of $(\il)^2$, $i$.$e$. beyond sector III, namely, in the
domain $\la^2>0,~k^2<0$, called sector I in \cite{shabad}. This is
the sector of bound states. Unlike sector III, in sector I the
Hilbert spaces, where  Hamiltonian $H$ (\ref{5}) and operator
$H^{III}$ (\ref{HIII}) act, consist of the same functions, since
one of the two solutions
$\left.\phi_{\pm\la,k}(r)\right|_{r\rightarrow 0}\asymp
r^{\pm\la+1/2}$ is ruled out as not belonging to
$L^2_\mu(0,\infty)$, the space of functions, square integrable
with the measure $\rmd\mu^{III}(r)$ (\ref{muIII}) on the interval
$(0,\infty)$: out of the two integrals \bee\label{ruleout}
\int_0\left|r^{\pm\la+\frac 1{2}}\right|^2\frac{\rmd
r}{r^2}=\int_0\frac{\rmd r}{r^{1\mp 2\la}}\eend one is and the
other is not equal to infinity.
 The $L^2$-solution
 satisfies the second boundary condition (\ref{bound}),
extended to the limit $r_L=0$, so we are within the same
eigenvalue problem (\ref{III}). On the other hand, the same
solutions are also ruled out by imposing artificial condition,
that the wave function should decrease fast enough in the origin,
- the procedure, accepted when the spectrum of bound states is
considered in physical text-books ( see $e.~g.$ \cite{QM} ). Thus,
the (finite number of) discrete states of $H$ and $H^{III}$ are
the same and can be presented in the form of  trajectories
$k=k_s(\la)$ or $\la=\la_s(k)$, $s=1,2,...s_0$.
 Thus the generalized Fourier expansion in the continuous limit
 may be written primarily for an arbitrary function $\widetilde{F}(r_*)\in
 L^2(-\infty,\infty)$
 in the $r_*$-representation as\bee\label{genfourier}
 \widetilde{F}(r_*)=\int_{-\infty}^\infty
 C(\la,k)\widetilde{\psi}_{\la,k}(r_*)\rmd\il+\sum_{s=1}^{s_0}
 C(\la_s(k),k)\widetilde{\psi}_{\la_s(k),k}(r_*),\nonumber\\
 C(\la,k)=\frac 1{r_0^2}\int_{-\infty}^\infty\widetilde{F}(r_*)\widetilde
 {\psi}^*_{\la,k}(r_*)\rmd r_*,\eend
 and secondary in the initial variable representation for arbitrary
 function $F(r)=(r/r_0)^{1/2}\widetilde{F}(r_*),~~F(r)\in L^2_\mu(0,
 \infty)$ as\bee\label{genfourier2}{F}(r)=\int_{-\infty}^\infty
 C(\la,k){\psi}_{\la,k}(r)\rmd\il+\sum_{s=1}^{s_0}
 C(\la_s(k),k){\psi}_{\la_s(k),k}(r),\nonumber\\
 C(\la,k)=\int_{0}^\infty{F}(r)
 {\psi}^*_{\la,k}(r)\frac{\rmd r}{r^2}.\eend

 It remains to comment on a generality of the boundary conditions
 (\ref{bound}). As the arbitrary dimensional parameter $r_0$
 involved in the transformation (\ref{21}), (\ref{22}) varies, the
 point $r_L=\rme^{-L/r_0}$, where the boundary condition
 (\ref{bound}) is imposed moves, provided that $L$ is fixed.
 This effectively changes the boundary conditions, considered in
 an unmoving point. Thus, the arbitrariness in $r_0$ reflects the
 arbitrariness in choosing self-adjoint boundary conditions, or in
 other words in fixing the self-adjoint extension. According to a
 general theorem \cite{naimark}, the spectrum in the continuum limit does not
 depend on this arbitrariness. What does depend, is the $S$-matrix
 (\ref{SIII}). The  kinematic unitary factor, containing $r_0$ in it, is
connected with the known $U(1)$ arbitrariness in fixing the
 self-adjoint extension.

\section{Two-channel sector}\label{Two-chan}
Now consider the domain of parameters $\la^2<0$, $k^2>0$, called
sector IV in ~\cite{shabad}. There, both fundamental solutions of
the \Sch are appropriate. The probability flux to/from the center
(from/to the infinity) may be nonzero. Correspondingly, the
boundary conditions should be of non-Sturm-Liouville type: they
should interconnect  values of the wave function, taken at the
opposite ends of the interval, like periodic or antiperiodic.

In this sector we take for definiteness $\il<0,~k>0$ throughout.

It is not adequate to try to extend equation (\ref{III}), or
(\ref{24}) beyond sector III into sector IV by including positive
$k^2$ into consideration. In that case equation (\ref{24}) would
correspond to negative, exponentially growing in absolute value
with $r\rightarrow\infty$, potential. Such a problem has a
discrete spectrum, unlimited from below ($cf$ the example
considered in Section 5.8 of the textbook \cite{titchmarsh} and
Appedix II of its Russian edition), similar to the energy spectrum
of \cite{case}, \cite{morse}, \cite{meetz}. Our study of sector IV
will be done using an operator  $H^{IV}$  (\ref{HIV}), coinciding
with $H^{III}$
 (\ref{HIII}) for small $r$ and with $H$ (\ref{5}) for large $r$.
\subsection{Kamke eigenvalue problem}
In sector IV define the differential operation $H^{IV}$
(\ref{HIV})
so that the \Sch  (\ref{4})  take the form
  \begin{equation}\label{IV}
H^{IV}\psi (r)=\left(k^2-\frac{\lambda^2}{r^2}\right)\psi (r).
\end{equation}
Let us introduce the new dimensional parameter $R$, real in sector
IV, according to the relation $R=-\il/k$. In what follows the
couple $\la,~R$ will be  used to parameterize the phase space
instead of the couple $\la~,k$. Then eq. (\ref{IV}) turns into
equation (\ref{x})
 We consider this equation in the box
\begin{equation}\label{box}
r_L\leq r\leq r_U.
\end{equation}
The lower limit of the box $r_L$ is meant to tend to zero, whereas
the upper limit $r_U$ to infinity \bee\label{limits} r_L=R{\rm
e}^{-\xi_L},\qquad r_U=R\xi_U. \eend  These limits  contain the
dependence on the ratio $R$ - but not on the eigenvalue $\il$, -
whereas $\xi_{L,U}$ are independent numbers, which will be taken
infinite later. The differential equation (\ref{x}), defined on
the interval (\ref{box}) and supplemented with necessary boundary
conditions make again the $general$ $eigenvalue$ $problem$ of
Kamke~\cite{Kamke}. In the case of interest here the choice of the
boundary conditions is restricted by the requirement that these
should survive the limiting process
$r_L\rightarrow0,~~r_U\rightarrow\infty$. This requirement is met,
for example, by the following  conditions imposed at the walls of
the box (\ref{box}) \bee\label{boundIV} \kern-50pt\frac{\psi
(r_L)}{(r_L)^\frac 1{2}}=\pm R^{-\frac 1{2}}\psi(r_U),\nonumber\\
(r_L)^{\frac 1{2}}\left.\frac{\rmd\psi(r)}{\rmd
r}\right|_{r=r_L}-\frac{\psi(r_L)}{2(r_L)^\frac 1{2}}
=\pm R^\frac 1{2}\left.\frac{\rmd\psi(r)}{\rmd
r}\right|_{~r=r_U}. \eend It is important, that the coefficients
in (\ref{boundIV}), as well as the limits $r_{L,R}$ (\ref{limits})
do not depend on the eigenvalue $\il$ but only contain the ratio
$R$. 
The matrix elements, defined as
\begin{equation}\label{matrix2}
H_{i j}^{IV} = \int_{r_L}^{r_U} \psi_{\la_i,R}^*(r) H^{IV}
\psi_{\la_j,R} (r)\rmd r,
\end{equation}
do satisfy the Hermiticity condition \bee\label{hermite2}
 (H^{IV}_{ji})^*-H^{IV}_{ij}\equiv
\left.\left(\psi_{\la_j, R}(r)\frac{\rmd\psi_{\la_i, R}^*(r)}
{\rmd r}-
      \psi_{\la_i, R}^*(r)\frac{\rmd\psi_{\la_{j}, R}(r)}{\rmd r}\right)\right|_{r_L}^{r_U}=0,
\eend once eqs.~(\ref{boundIV}) are fulfilled for each of the
functions $\psi_{\la_{i,j}, R}(r)$. The origin of the boundary
conditions (\ref{boundIV}) will become clear
 below in this section.
 The special choice (\ref{limits}) of
dependence of $r_{L,U}$ on $R$ is not important for providing the
Hermiticity.

Certainly, the boundary conditions (\ref{boundIV}) are not the
most general conditions, meeting the above requirements. A more
general choice might be provided, if one introduced an arbitrary
parameter with the dimensionality of length $r_0$ in place of
$R=-\il/k$ in (\ref{boundIV}). This would yield unreasonable
complications in handling the spectra and eigenfunctions without,
however, affecting the important conclusions about the
condensation of eigenvalues into the continuum in the limit
$r_L=0,~R_U=\infty$. Therefore, unlike the previous treatment in
Section $\bf 2$, we do not keep arbitrary $r_0$ in this Section.
Its possible effect for the scattering matrix will be discussed in
Subsection $\bf 3.3$.

The Hermiticity condition (\ref{hermite2}), when taken with $i=j$,
reads that the probability flux (\ref{27a}) is the same at the two
opposite walls of  the box (\ref{box}) . Thus, the boundary
conditions (\ref{boundIV}) agree with the probability
conservation, but, unlike the boundary conditions (\ref{bound})
imposed in sector III, admit that the probability flux (\ref{27a})
be nonzero. This means that the overall probability may flow
either into or out of the system, depending upon the solution
selected.

Following the spectral theory \cite{Kamke}, \cite{naimark}, we
conclude that the special eigenvalue problem (\ref{x}),
(\ref{boundIV}) should for every $R$ have two countable manifolds
of infinitely growing eigenvalues $(\il_m(R))^2$ and ${\rm
Im}(\overline{\la}_m(R))^2$, $m=0,1,2...$, which
alternate:\bee\label{alternate} (\il_m)^2\leq {\rm
Im}(\overline{\la}_m)^2<(\il_{m+1})^2.\eend (The lowest value
$({\rm Im}\overline{\la}_0)^2$ only exists for the upper sign in
(\ref{boundIV}).) Thus, we face a discrete spectrum, as long as
the box wall positions (\ref{box}) are finite,
 $r_L\neq 0, ~r_U\neq\infty$. The spectrum is constituted by discrete
trajectories $\il_n(k)$ labelled by the integer $n$. The
trajectories are expected to condense to form a continuum of
states, when $r_L\rightarrow 0$ and $r_U\rightarrow\infty$. The
spectral theory predicts that, at least for large $(\il)^2$, the
spacings between neighboring eigenvalues within one manifold are
\bee\label{spacing} \il_{n+1}-\il_n=\frac{2\pi}{N},\quad N=
\int_{r_L}^{r_U}\sqrt{ \frac{1}{R^2}+\frac 1{r^2}}~\rmd r,\eend
and the same for ${\rm Im}\overline{\la}_n$. The integral $N$ here
plays the role of the size of the box. It diverges both at the
lower and the upper limits as $r_L\rightarrow 0$ and
$r_U\rightarrow\infty$,~~$N=\xi_L+\xi_U$,~~ thus providing the
vanishing of the spacings . We shall study the spectrum
specifically in this limiting case of interest  in the next
subsection to see that the spectral trajectories do condense
everywhere throughout sector IV, not only for large $(\il)^2$.

 By reducing the spectral problem (\ref{x}), (\ref{boundIV}) to
 the standard Liouville form we shall explicitly see below that its
eigenfunctions are a complete set for every $R$. If all the
eigenvalues in (\ref{x}) are positive, $i.e.$ belong to sector IV,
the complete set in the limiting case is exhausted by the
functions belonging to the continuum. If there  are also several
negative eigenvalues $\il^2_s$ (real $\la_n$), this means that
there exist usual bound states, since, with $R$ fixed, the
corresponding values of $k$ become imaginary, and we enter the
sector of bound states (called sector I in \cite{shabad}) along
the ray $(-\il/k)=R=const$. The discussion of this point presented
in Section $\bf 2$, might be repeated here, with the only
reservation that there we entered sector I along the ray ${\rm
Im}k=const$. The corresponding (finite number of) eigenfunctions
make the complete set when taken together with the eigenfunctions,
which belong to the continuum.

 The analog of
Eq.~(\ref{orthog}) is the following orthogonality relation
 in sector IV
\begin{equation}\label{orthogIV}
\int_{r_L}^{r_U}\psi^*_{\la,R}(r)\psi_{\la',R'}(r)
\left((k')^2-k^2+\frac{(\il)^2-(\il')^2}{r^2}\right)\rmd r = 0.
\end{equation}
Any two solutions that belong to the same ray in the
($\il,k$)-plane ($i. e.$ have common value of $R=-\il/k$) are
mutually orthogonal with the universal measure (\ref{measureIV})
 provided that
the eigenvalues $(\il)^2$ do not coincide,
$((\il)^2\neq(\il')^2)$:
\begin{equation}\label{orthogIVu}
((\il)^2-(\il')^2)\int_{r_L}^{r_U}\psi^*_{\la,R}(r)\psi_{\la',R}(r)
\left(\frac{1}{R^2}+\frac{1}{r^2}\right)\rmd r = 0 . 
\end{equation}

%
%
%
\subsection{Spectrum}
In sector IV every solution of the \Sch is meaningful, since it
oscillates at the both ends of the interval. ( We exclude the
value $\la=0$, which requires a special treatment to be done below
in this Subsection.) Let $\phi_{\pm\la,k}(r)$ be the solutions,
that behave in the origin like eq.~(\ref{1.2}): \bee\label{phi1}
\phi_{\pm\la,k}(r)\asymp r^{\pm\rmi\rm{Im}\lambda+\frac 1{2}}.
\eend These are expressed in sector IV as \cite{regge}
\begin{eqnarray}\label{c}
\p=E\f+G\fk,\nonumber\\
\pmin=G^*\f+E^*\fk
\end{eqnarray}
in terms of the solutions $f_{\pm \la ,k}(r)$ that behave like
$\exp (\mp\rmi kr)$ at infinity: \bee\label{f1} f_{\la,\pm k}(r
)\asymp\exp(\mp\rmi kr). \eend The constants  $E,~G$  are
connected with the Jost functions (\ref{11}) as \bee\label{GE}
f(\la,k)=2ikG,   \quad    f(\la,-k)=-2ikE \eend and are in sector
IV subject to the relation \cite{regge} (we corrected the obvious
dimension-violating misprint in eq.(5.13) of (\cite{regge}))
\bee\label{defect}|E|^2-|G|^2=-\frac\il{k}\equiv R. \eend

Let us look for solution to the problem (\ref{x}), (\ref{boundIV})
in the form of a linear combination of fundamental solutions of
the differential equation  (\ref{HIV}) \bee\label{psi1}
a\p+b\pmin. \eend We restrict ourselves to  the case when $\il$
and $k$ are of opposite signs, $R>0$. Using (\ref{c}) and the
asymptotic forms (\ref{phi1}),  (\ref{f1}) one writes the boundary
conditions (\ref{boundIV}) in the form of the set of equations for
the coefficients $a,~b$ in (\ref{psi1}), valid provided that $r_L$
is much less, while ~$r_U$ is much greater than all dimensional
parameters in the problem, $i.e.$ than $R$ and other dimensional
parameters, on which the potential $V(r)$ may depend: \bee
\kern+50pt a\left(\pm R^{1/2} r_L^{\rmi\il}-E\exp(-\rmi k r_U)-
 G \exp(\rmi kr_R)\right)\hspace{25cm}
\nonumber\\
 +b\left(\pm R^{1/2}r_L^{-\rmi\il}-E^*\exp(\rmi k r_U)-G^*\exp(-\rmi k
r_U)\right) =0, \hspace{25cm}\nonumber\\a\left(\mp R^{1/2}
r_L^{\rmi\il}+ E\exp(-\rmi k r_U)-
G\exp(\rmi k r_U)\right)\nonumber\hspace{25cm}\\
 -b\left(\mp R^{1/2}
r_L^{-\rmi\il}+E^*\exp(\rmi k r_U)- G^*\exp(-\rmi k r_U)\right)
=0.\hspace{25cm} \nonumber \eend \bee\label{set} \eend This set is
simplified by linearly combining  the equations as follows:
\bee\label{set1} \kern+50pt aG\exp(\rmi k r_U)+b(\mp R^{1/2}
r_L^{-\rmi\il}+E^*\exp(\rmi k r_U))=0\nonumber\\
\vspace{20cm}
 a(\mp R^{1/2}
r_L^{\rmi\il}+ E\exp(-\rmi k r_U))+ bG^*\exp(-\rmi k r_U)=0.\eend
The spectrum is obtained by equalizing the determinant of this set
with zero. Using (\ref{defect}) one gets

\bee\label{det}  (Ew_-+E^*w_-^*) =\pm 2R^{1/2}, \eend where
\begin{equation}\label{w} w_-=r_L^{-\rmi\il}~{\rm e}^{-\rmi k
r_U}=R^{-\rmi\il}~{\rm e}^{\rmi\il(\xi_L+\xi_U)}.
\end{equation}

Define the three real angles $\delta_{1,2},~\alpha$ (we shall need
$\delta_1$ later) \bee\label{62} \delta_1=\arg f(\la,k),
\hspace{10mm}
\delta_2=\arg f(\la,-k), \nonumber\\
\cos\alpha=\left|\frac{f(\la,k)}{f(\la,-k)}\right|,\quad
\sin\alpha= \frac{2\il}{R^{1/2}|f(\la,-k)|}, \quad
-\frac\pi{2}\leq \alpha\leq\frac\pi{2}.\eend Eq.(\ref{defect})~
guarantees the fulfillment of the necessary equality
$\sin^2\alpha+\cos^2\alpha=1$.
 Bearing in mind (\ref{GE}), (\ref{defect}) we write equation (\ref{det}) in the
form \bee\label{det2} \pm \sin\alpha -\sin\left(\delta_2 -\il~\ln
R +\il(\xi_L+\xi_U)\right)=0.\hspace{2cm} \eend This equation
reduces to two infinite series of equations, wherein $\alpha$ and
$\delta_2$ are functions  of $R$ and of $\il_n$, the latter being
set equal to $\il^{(1)}_n$ in the first, and to $\il^{(2)}_n$ in
the second equation:
 \bee\label{prefinal}
 \il^{(1)}_n=\frac{2n\pi}{\xi_L+\xi_U} +\frac {1}{\xi_L+\xi_U}\left(
 \pm\alpha-\delta_2+\il^{(1)}_n\ln R\right),\nonumber\\
\il^{(2)}_n=\frac{(2n+1)\pi}{\xi_L+\xi_U} +\frac
{1}{\xi_L+\xi_U}\left(
 \mp\alpha-\delta_2+{\rm Im}\la^{(2)}_n\ln R\right),\nonumber\\
 \quad n=0,\pm 1,\pm 2 ... .\eend Remind that the double sign
 here corresponds to that in
 (\ref{boundIV}). From these the eigenvalues
$\il^{(1,2)}_n$
are to be found for each $n$. This result agrees with
(\ref{spacing}).

Finally, the asymptotic form of the discrete spectrum of the
boundary problem (\ref{IV}),~(\ref{boundIV}) in the limit
$r_L\rightarrow0,~r_U\rightarrow\infty $ in sector IV is 
 given as \bee\label{final} \il^{(1)}_n=\frac{2n\pi}{\xi_L+\xi_U},\quad
 \il^{(2)}_n=\frac{(2n+1)\pi}{\xi_L+\xi_U}.\eend This does not depend on details
of the interaction $V(r)$ and on the choice of the sign in
(\ref{boundIV}).

For the further analysis we shall need  the relations
\cite{regge}, valid in sector IV:\bee\label{complex}
\phi^*_{\la,k}(r)=\pmin,\qquad
f^{*}_{\la,k}(r)=\fk,\nonumber\\
f^*(\la,k)=f(-\la,-k), \eend  supplemented with the relations
\bee\label{minus} \p=\pk,\quad\f=\fmin,\eend which are a
consequence of the asymptotic behavior (\ref{phi1}),~(\ref{f1})
and the evenness of (\ref{IV}) with respect to reflection of $k$
or $\la$.

Let us introduce the common enumeration of the eigenvalues
(\ref{prefinal}), (\ref{final})\bee\label{enumeration}
\il_m=\il^{(1)}_{m/2}\quad {\rm for}~ m~{\rm even},\nonumber\\
\il_m=\il^{(2)}_{(m-1)/2}\quad {\rm for}~ m~{\rm odd}.\eend From
the second line of (\ref{complex}) it follows that $~\alpha~$
 and $\delta_2~$ in (\ref{62}) are odd with respect
to $\il$ (keeping $R$ invariant). This implies that the
eigenvalues $\il^{(1,2)}$,  defined as solutions of equations
(\ref{prefinal}), obey the relations:\bee\label{il}
\il_n^{(1)}=-\il_{-n}^{(1)},\qquad
\il_n^{(2)}=-\il_{-1-n}^{(2)}\nonumber\\
{\rm or}\nonumber\\
\il_m=-\il_{-m}.\eend This means that each eigenvalue
$(\il_m)^2\neq 0$ of the operator $H^{IV}$, eq.~(\ref{HIV}), is
two-fold degenerate. 
If one identifies $\il_{-m}$ with
Im$\bar\la_m$, introduced in Subsection $\bf 3.1$, one sees that
this degeneracy
 corresponds to the equality signs in the chain of weak
 inequalities (\ref{alternate}).

Eqs.(\ref{il}) do not hold true for $m=0$. Indeed, eqs.(\ref{il})
would imply, that $\il_0\equiv\il^{(1)}_0=0$. This is not the
case, however: the oddness of $\alpha$ and $\delta_2$ does not yet
provide that $\il^{(1)}_0=0$ be a solution of equation
(\ref{prefinal}), since $\delta_2=\arg f(\la,-k)$ in it is,
generally, not a continuous function in the point $\la=0, R\neq
0$. For instance, in the free case $V=0$, when the Jost function
 is known, it can be written in sector IV as:
\bee\label{f^0}
f^{(0)}(\la,-k)=\left(\frac\la{2R}\right)^{-\la+\frac 1{2}}\Gamma
(1+\la)=\left(\frac{|\la|}{2R}\right)^{\frac 1{2}}|\Gamma
(1+\la)|~\exp\left(|\la|\frac\pi{2}+\rmi\delta^{(0)}_2\right),\nonumber\\
\delta^{(0)}_2=\arg f^{(0)}(\la,-k)=\frac\pi{4}~{\rm sgn}(\il)-
\il\ln\frac{|\il|}{2R}+\arg\Gamma(1+\rmi\il),\qquad\eend where
$\Gamma$ is the Euler gamma-function, and ${\rm sgn}(\il)$ is 1
for positive and -1 for negative arguments. Here the phase
$\delta^{(0)}_2$ contains the discontinuity ${\rm sgn}(\il)$, and
hence $\arg f^{(0)}(\la,-k)$ is not defined in the point $\il=0$.
Correspondingly, the boundary problem (\ref{x}), (\ref{boundIV})
with $V=0$ in (\ref{HIV}) has no solutions for $\la=0$, $i.e.$ the
boundary conditions (\ref{boundIV}) cannot be satisfied by
combining the fundamental solutions of the differential equation
(\ref{x}), which in this case are $\sqrt r$ and $\sqrt r\ln r.$
The said does not rule out the possibility that the point $\la=0$
might belong to the spectrum. This may happen for dynamical
reasons, for some $V(r)$. The statement above only means that the
general consideration alone are not enough for establishing the
existence of the zero mode.

It can be demonstrated that any other self-adjoint boundary
conditions used in place of (\ref{boundIV}) would lead to the same
result. The most important conclusion about the spectrum
(\ref{final}) is: in the domain of interest $r\in (0,\infty)$ the
spectral trajectories condense to make a continuum and to densely
cover the space of quantum numbers $(\il,~k)$ of sector IV.

\subsection{Orthonormal solutions}
To obtain the wave functions corresponding to the eigenvalues of
the limiting spectral problem found in the previous Subsection,
consider the first equation in (\ref{set1}). When taken on
solutions of equations (\ref{prefinal}), it becomes

\bee\label{combined1} G~a +(E^*\mp \varepsilon^{(j)}\rme^{\rmi
(\pm\varepsilon^{(j)}\alpha-\delta_2)}R^{1/2})~b=0,\quad j=1,2,
\eend where the factor $\varepsilon$ takes two different values:
$\varepsilon^{(1)}=1,~\varepsilon^{(2)}=-1$ respective to whether
the first or the second equation in (\ref{prefinal}) is used.
According to (\ref{enumeration}), $\varepsilon=(-1)^m$. The moduli
of the complex coefficients in front of  $a$ and $b$ in this
equation are the same\bee\label{modulus} | E\mp\varepsilon^{(j)}
\rme^{-\rmi
(\pm\varepsilon^{(j)}\alpha-\delta_2)}R^{1/2}|=|G|,\eend while
their phases are expressed as follows\bee\label{phases}\arg
G=\frac \pi{2}~{\rm
sgn}(\il)+\delta_1,\nonumber\\
\arg (E^*\mp\varepsilon^{(j)}\rme^{\rmi
(\pm\varepsilon^{(j)}\alpha-\delta_2)}R^{1/2})=\pm\varepsilon^{(j)}
\alpha-\delta_2+\frac \pi{2}~{\rm sgn}(\il).\eend Relations
(\ref{modulus}), (\ref{phases}) are direct consequences of the
definitions (\ref{GE}), (\ref{62}). To derive them,
eq.(\ref{defect}) and the relation\bee\label{devil1}
E^*\rme^{-\rmi
(\pm\varepsilon^{(j)}\alpha-\delta_2)}
=\pm\varepsilon^{(j)} R^{1/2}+\rmi\frac{|f(\la,k)|}{2\il}, \eend
are useful.

The two corresponding series of eigenfunctions (\ref{psi1}) have
the form\bee\label{eigenfunc}
\psi^{(j)}_{\la,R}(r)\equiv\psi_{\la^{(j)},R}(r)=\nonumber\\
=\frac{\rmi}{2\sqrt\pi}\left(\rme^{-\rmi(\mp\varepsilon^{(j)}
\alpha+\delta_1+\delta_2)/2}\phi_{\la,k}(r)-
\rme^{\rmi(\mp\varepsilon^{(j)}\alpha+\delta_1
+\delta_2)/2}\phi_{-\la,k}(r)\right),\nonumber\\
k=-\il/R. \eend It is understood, that $\il$ in
$\psi^{(j)}_{\la,R}(r)$ is $\il^{(j)}_n$, the solution of the
first $(j=1)$ and the second (j=2) equation in (\ref{prefinal}).
In the continuum limit the two series $\psi^{(1,2)}_{\la,R}(r)$
only differ due to the factor $\varepsilon^{(1,2)}$ in them. The
common phase factor has been chosen in such a way that the reality
of the eigenfunction (\ref{eigenfunc}) be provided. This follows
from the oddness of the angles $\alpha,~\delta_{1,2}$ (\ref{62})
under reflection of sign of $\la$, with $R$ kept constant. The
latter property is proved using (\ref{complex}) and (\ref{minus}).
The oddness of $\alpha,~\delta_{1,2}$ also leads to that of the
eigenfunctions (\ref{eigenfunc}):\bee\label{reality}
(\psi^{(j)}_{\la,R}(r))^*=\psi^{(j)}_{\la,R}(r)=-\psi^{(j)}_{-\la,R}(r).
\eend

 Note the
important difference with the standard Fourier analysis, based on
the eigenvalue problem ~~$-$d$^2y$/d$\xi^2=p^2y$,~~ $y(-\xi_L)=
y(\xi_U)$,~~$y'(-\xi_L)= y'(\xi_U)~~$, where there are two
independent eigenfunctions $\exp (\pm\rmi\p\xi)$ ~~( connected by
the complex conjugation operation), related to the same eigenvalue
$p^2$. In that case the corresponding coefficients $a$ and $b$ are
not subject to an equation like (\ref{combined1}), and remain
arbitrary. On the contrary, in our case the complex conjugation
operation, when applied to an arbitrarily normalized
eigenfunction, $does$ $not$ create any new, independent solution
of the boundary problem, but only multiplies it by a unit complex
factor. This explains why there is only one, sine-like,
eigenfunction (\ref{eigenfunc}), whereas the other, cosine-like,
eigenfunction is absent.

The probability flux (\ref{27a}) calculated with the wave function
(\ref{eigenfunc}) is zero. Certainly, a linear combination of
eigenfunctions with complex coefficients carries nonzero flux to
or from the center. In this respect the situation is different
from the bound states, the confined states of sector III, or from
the elastic scattering states in what is called sector II in
\cite{shabad}, where not only the eigenfunctions do not carry
probability flux, but any their linear combinations do not either,
since all the wave functions disappear at the both ends of the
box, in accord with the Sturm-Liouville boundary conditions, used
in these sectors. Remind, that the self-adjoint boundary
conditions (\ref{boundIV}) are not of the Sturm-Liouville type.

With the use of (\ref{c}) and of definitions (\ref{62}) the same
eigenfunction (\ref{eigenfunc}) may be also presented as
\bee\label{eigenfunc1}
\psi^{(j)}_{\la,R}(r)=\nonumber\\=\frac{\pm\rmi\varepsilon^{(j)}}{2}
\left(\frac{R} {\pi}\right)^{1/2} \left(\rme^{\rmi
(\mp\varepsilon^{(j)}\alpha-\delta_1+\delta_2)/2}f_{\la,k}(r)-
\rme^{-\rmi
(\mp\varepsilon^{(j)}\alpha-\delta_1+\delta_2)/2}~f_{\la,-k}(r)\right),
\nonumber\\
 k=-\il/R. \eend
The eigenfunction (\ref{eigenfunc}), (\ref{eigenfunc1})  behaves
near $r=0$ as\bee\label{1r=0} \psi^{(j)}_{\la,R}(r)\asymp
\left(\frac r{\pi}\right)^{1/2}\sin
\left(\frac{\mp\varepsilon^{(j)}\alpha+\delta_1 +\delta_2}{2}-\ln
r~\il\right) \eend and near $r=\infty$ as\bee\label{1r=infty}
\psi^{(j)}_{\la,R}(r)\asymp \mp\varepsilon^{(j)}\left(\frac
R{\pi}\right)^{1/2}\sin
\left(\frac{\mp\varepsilon^{(j)}\alpha-\delta_1 +\delta_2}{2}-kr\right),
\qquad k=-\il/R.\eend These are standing waves, that do not vanish
at any of the end points, unlike the standing wave (\ref{near=0})
in sector III, which vanishes ar $r=r_L$, or the standing wave,
corresponding to the usual elastic scattering in sector II , which
vanishes at the remote end of the box $r=r_U$.

The eigensolutions of the self-adjoint boundary problem under
consideration in sector IV in the quadrant sgn$(k\il)=-1$ , taken
in any of the forms (\ref{eigenfunc}) or (\ref{eigenfunc1}),
satisfy the following orthonormality relation (the scalar product
with the measure (\ref{measureIV})) in the asymptotic regime
$\xi_L,~\xi_U~\rightarrow\infty$ ($r_L\rightarrow
0,~r_U~\rightarrow\infty$)
\bee\label{ortnorm}\left(\psi^{(j)}_{\la,R}(r),
\psi^{(i)}_{\la',R}(r)\right)\equiv
\int_{r_L}^{r_U}(\psi^{(j)}_{\la,R}(r))^*\psi^{(i)}_{\la',R}(r)
\left(\frac{1}{R^2}+\frac{1}{r^2}\right)\rmd
r=\nonumber\\=\delta_{ij}~\delta_{nn'}~\frac{\xi_L+\xi_U}{2\pi},\quad
i,j=1,2, \quad n=1,2,3... .\eend In the limit $\xi_L=\xi_U=\infty$
this is \bee\label{ortnorm1}\left(\psi^{(j)}_{\la,R}(r),
\psi^{(i)}_{\la',R}(r)\right)\equiv
\int_0^{\infty}(\psi^{(j)}_{\la,R}(r))^*\psi^{(i)}_{\la,R}(r)
\left(\frac{1}{R^2}+\frac{1}{r^2}\right)\rmd r=\nonumber\\
=\delta_{ij}~ \delta(\il-\il'). \eend The complex conjugation sign
may be omitted here, since the eigenfunctions
$\psi^{(j)}_{\la,R}(r)$ have been chosen real. To see, that
eqs.(\ref{ortnorm}), (\ref{ortnorm1}) hold true, we follow the
standard procedure \cite{QM}. First note, that the scalar product
of eigenfunctions with $n\neq n'$, $i\neq j$ is zero according to
(\ref{orthogIVu}). Then, we only need to calculate the
contributions into (\ref{ortnorm}), (\ref{ortnorm1}), originating
from the coinciding values $n=n'$, $i=j$. These are divergent due
to integration near the end points. To find these contributions,
the asymptotic expressions (\ref{1r=0}), (\ref{1r=infty}) are
sufficient. When integrating near $r=r_L$, we take (\ref{1r=0}),
neglect $1/R^2$ as compared to $1/r^2$ in the measure, and use the
new integration variable $\xi=\ln (r/R)$. When integrating near
$r=r_U$, we take (\ref{1r=infty}), neglect $1/r^2$ as compared
with $1/R^2$ in the measure, and use the new integration variable
$\xi=Rr$. In this way  integral (\ref{ortnorm}) is reduced to
$(1/2\pi)\int_{-\xi_L}^{\xi_U}\exp\{\rmi(\il'-\il)\}\rmd\xi$,
where the limits are given as (\ref{limits}).

\subsection{Generalized Fourier expansion}
According to the spectral theory \cite{Kamke},\cite{naimark}, the
sets of eigenfunctions associated with the self-adjoint boundary
problem (\ref{x}), (\ref{boundIV}) should be complete. We shall
see this directly later, after we reduce Eq. (\ref{IV}) to the
normal Liouville form. Now, using the family of orthonormalized
eigenfunctions (\ref{eigenfunc}) or, which is the same,
(\ref{eigenfunc1})
 found in the previous subsection for the case sgn($\il~k)=-$1,
 we may write the Fourier
expansion of a sufficiently smooth arbitrary function $F(r)$ in
the Hilbert space of functions with finite norm
$\int_0^\infty|f(r)|^2\rmd\mu_{IV}(r)<\infty$, under the
assumption that there are no bound states in the problem, $i$.$e$.
that the eigenvalues $(\il)^2$ of $H^{IV}$ eq.(\ref{x}) are all
positive. Negative eigenvalues, if they exist, are not covered by
the analysis above, since they fall out of sector IV and the
corresponding fundamental solutions do not behave like
(\ref{phi1}), (\ref {f1}), contrary to what was assumed in
Subsection $\bf 3.2$. 

In the asymptotic regime when the upper position $r_U$ of the box
wall tends to infinity, while its lower position $r_L$ tends to
zero, this expansion is :\bee\label{Fourier}
F(r)=\frac{2\pi}{\xi_L+\xi_U} \sum_{j=1,2}\sum_{~n=1}^{\infty}
C(\la_n^{(j)},R)\psi_{\la_n^{(j)},R}(r).\eend In the continuum
limit $r_L=0$, $r_U=\infty$ this
becomes\bee\label{fouriercont}F(r)=\sum_{j=1,2}\int_0^{\infty}
C^{(j)}(\la,R)\psi_{\la,R}^{(j)}\rmd\il.\eend We introduced
d$\il\equiv \il^{(j)}_{n+1}-\il^{(j)}_n=2\pi/(\xi_L+\xi_U)$ in
accord with (\ref{final}). It is meant throughout, that $R=-\il/k$
is positive and kept constant while summing or integrating over
$\il$. The expansion coefficients follow from (\ref{ortnorm}),
(\ref{ortnorm1}) to be in (\ref{Fourier})
\bee\label{coefficients}C(\la_n^{(j)},R)=\int_{r_L}^{r_U}(\psi_{\la^{(j)}_n,R}(r))^*
F(r)\rmd\mu(r)\nonumber\\
\eend and in (\ref{fouriercont})
\bee\label{coefficients1}C^{(j)}(\la,R)=\int_0^{\infty}(\psi^{(j)}_{\la,R}(r))^*
F(r)\rmd\mu(r)\nonumber\\
\eend The measure  d$\mu (r)$ is defined as (\ref{measureIV}). We
should have marked $C$, as well as $\psi$, here with the label
$\pm$ to indicate the dependence of these quantities upon the
choice of sign in the boundary conditions (\ref{boundIV}), which
we did not, however, to avoid an excessive complexity of
notations.

For convergence of (\ref{coefficients}), (\ref{coefficients1}) it
is needed that \bee\label{converge} {\rm lim}~_{r\rightarrow
0}\frac{F(r)}{\sqrt r}=0,\quad F(\infty)=0.\eend

The expansion coefficients $C$, initially defined for $n\geq 1$,
$\il>0$, can be extended to negative values of $n$ or $\il$, using
(\ref{reality}), and to $\la_0=0$ as \bee\label{extend}
C(\la_{-n}^{(j)},R)=C(-\la_n^{(j)},R)=-C(\la_n^{(j)},R)\nonumber\\
C^{(j)}(-\la,R)=-C^{(j)}(\la,R),\quad C(0)=0.\eend Then, with the
use of (\ref{eigenfunc}), or (\ref{eigenfunc1}), we can reduce the
direct (\ref{Fourier}) and inverse (\ref{fouriercont}) Fourier
transformations to a transformation, referring to the fundamental
solutions $\phi_{\la,k}(r)$ and $\f_{\la,k}(r)$. One has
\bee\label{C}F(r)=\frac{\rmi \sqrt\pi}{\xi_L+\xi_U}
\sum_{j=1,2}\sum_{n=-\infty}^{\infty}
C(\la_n^{(j)},R)\rme^{-\rmi(\mp\varepsilon^{(j)}
\alpha+\delta_1+\delta_2)/2}\phi_{\la,k}(r)=\nonumber\\
=\frac{(\pi R)^{1/2}}{\xi_L+\xi_U}
\sum_{j=1,2}\sum_{n=-\infty}^{\infty}\mp\rmi\varepsilon^{(j)}
C(\la_n^{(j)},R)\rme^{\rmi(\mp\varepsilon^{(j)}
\alpha-\delta_1+\delta_2)/2}f_{\la,k}(r) \eend in the asymptotic
regime, and
 \bee\label{Ccon}
F(r)=\frac{\rmi}{2\sqrt\pi}\sum_{j=1,2}\int_{-\infty}^{\infty}
C^{(j)}(\la,R)\rme^{-\rmi(\mp\varepsilon^{(j)}
\alpha+\delta_1+\delta_2)/2}\phi_{\la,k}(r)\rmd\la=\nonumber\\
=\frac{\mp\rmi}{2}\left(\frac R{\pi}\right)^{1/2}
\sum_{j=1,2}\varepsilon^{(j)}\int_{-\infty}^{\infty}
C^{(j)}(\la,R)\rme^{\rmi(\mp\varepsilon^{(j)}
\alpha-\delta_1+\delta_2)/2}f_{\la,k}(r)\rmd\la\eend in the
continuum limit.

Writing the latter expansion in the form of a transformation with
respect to the fundamental solution $\phi_{\la,k}(r)$:
\bee\label{simple} F(r)=\int_{-\infty}^{\infty}
D(\la,R)\phi_{\la,k}(r)\rmd\il,\quad k=-\frac\il{R},
\eend where\bee\label{D} D(\la,R)=\rmi\frac{
\sqrt\pi}{2}\rme^{-\rmi\frac{\delta_1+\delta_2}{2}}\sum_{j=1,2}
C^{(j)}(\la,R)\rme^{\pm\rmi\varepsilon^{(j)}\frac{\alpha}{2}}=\nonumber\\
=-\frac 1{2}
\int_0^{\infty}\left(\phi_{\la,k}(r)\cos\alpha~\rme^{-\rmi(\delta_1+\delta_2)}
-\phi_{-\la,k}(r)\right)F(r)\rmd\mu(r),\eend with the help of
(\ref{phases}) we finally obtain  the
 transformation, inverse to (\ref{simple}),\bee\label{Dla}
D(\la,R)=\frac{\rmi\il}{f(\la,-k)}\int_0^{\infty}
F(r)f_{\la,-k}(r)\rmd\mu(r),\quad k=-\frac\il{R},\eend wherein
$f_{\la,-k}(r)$ is the other, independent fundamental solution.

We can also write a transformation, dual to (\ref{simple}),
(\ref{Dla}). Writing the second line of (\ref{Ccon}) in the form
\bee\label{simple1} F(r)=\int_{-\infty}^\infty
B(\la,R)\f\rmd\il,\quad k=-\frac\il{R},\eend where\bee\label{B}
B(\la,R)=\frac{\pm \rmi}{2}\left(\frac R{\pi}\right)^{\frac
1{2}}\left(\sum_{j=1,2}\varepsilon^{(j)}\rme^{\mp\varepsilon^{(j)}
\alpha/2}C^{(j)}(\la,R)\right)\rme^{\rmi
(\delta_2-\delta_1)/2}=\nonumber\\=-\rmi\frac{\sqrt
R}{2\pi}\rme^{\rmi\delta_2}\sin\alpha\int_0^\infty\ F(r)
\phi_{-\la,k}(r)\rmd\mu (r). \eend We used (\ref{eigenfunc}) in
the expression for $C$ (\ref{coefficients1}). With the help of
(\ref{phases}) we finally obtain the transformation, inverse to
(\ref{simple1}) in the
form\bee\label{Bla}B(\la,R)=\frac{-\rmi\il}{\pi
f(-\la,k)}\int_0^\infty F(r)\phi_{-\la,k}(r)\rmd\mu (r),\quad
k=-\frac\il{R}. \eend

It is remarkable, that the transforms $D(\la,R)$ and $B(\la,R)$ do
not depend upon choice of sign in (\ref{boundIV}), in other words
the validity of transformations (\ref{simple}), (\ref{Dla}) and
(\ref{simple1}), (\ref{Bla}) could be established with the help of
any of the two self-adjoint limiting boundary problems.

 If there are
 some bound states,  after equation (\ref{x}) is continued beyond
 sector IV,
 the expansion (\ref{Fourier}) should
 be supplemented by the sum, with $R$ being the same positive
 parameter as in the rest of the expansion (\ref{Fourier}),
 \bee\label{suppl}
 \sum_{s=1}^{s_0}C_s\phi_{\la_s,k_s}(r),\quad {\rm
 Im}k_s=\frac{{\rm Re}\la_s}{R}>0,\quad {\rm Re}k=\il=0\eend
 where\bee\label{phif}\phi_{\la_s,k_s}(r)=\frac{-R}{2\la_s}f(\la_s,k_s)f_{\la_s,-k_s}(r)\eend
 is the solution of equation (\ref{x}), taken on the zeros of the
 Jost function $f(\la_s,-k_s)=0$ (this means that $E=0$ in
 (\ref{c})). It decreases, when $r\rightarrow 0$, as
 $r^{-\la+1/2}$, and as $\exp \{-{\rm Im}k ~r\}$, when
 $r\rightarrow\infty.$ Hence, the both sides in eqs.(\ref{boundIV}),
  expressing  the boundary conditions,  vanish in the limit $r_L=0$,
   $r_U=\infty$, and
  these conditions are satisfied. Correspondingly, two
  eigenfunctions $\phi_{\la_s,k_s}(r)$ are orthogonal with the measure
   (\ref{measureIV}),
  provided that the values of $\la_s$ for them are different, but
  the values of $R={\rm Re}\la_s/{\rm Im}k_s$ are coinciding.
  Therefore, (\ref{phif}) make solutions to the eigenvalue problem
  (\ref{x}), (\ref{boundIV}).

\subsection{S-matrix}
 Restrict ourselves to the
case $\il <0$, Re$k>0$ for definiteness. By equalizing the
probability fluxes near the origin and the infinitely remote point
we get from ~(\ref{b}) and ~(\ref{c}) in a standard way the
probability conservation relations \bee\label{52}
 \frac{|G|^2}{|E|^2}-\frac{{\rm Im}\la}{k|E|^2}=1,    \qquad
\frac{|C|^2}{|D|^2}-\frac k{{\rm Im}\la|D|^2}=1. \eend The first
eq.~(\ref{52}) means that the coefficient of reflection of the
wave,

 incoming
from infinity, plus the coefficient of transmission of this wave
into the inner world makes  unity. The second eq.~(\ref{52})
reads: the coefficient of reflection of the wave, emitted from the
origin,
 back into the origin
plus the coefficient of transmission of the emitted wave to
infinity is unity.

  Define the two two-component columns
\bee\label{58} \kern-50pt\Lambda=\left(\begin{array}{c}
\fk\vspace{7mm}\\\sigma\p\end{array}\right),  \quad
\Lambda'=\left(\begin{array}{c}
\f\vspace{7mm}\\\sigma\phi_{-\la,k}(r)\end{array}\right),\quad
\rm{with}\quad \sigma=\left(\frac k{\il}\right)^\frac 1{2}. \eend
Then relations (\ref{b}) and the first one of (\ref{c}) can be
expressed using a square $2\times 2$ matrix $S$ \bee\label{S}
\Lambda'=S\Lambda. \eend The unitarity of the $S$-matrix
\bee\label{59} SS^\dagger=1, \eend where $S^\dagger$ is Hermitian
conjugate to $S$, follows from the  relations  (\ref{52}) and from
the complex conjugation rules ~\cite{regge} for the Jost
functions:
\begin{equation}\label{15}
f^*(\lambda,k)=f(\lambda^*,k^*\exp(-\rmi \pi)).
\end{equation}
The $S$-matrix in~(\ref{S}) is (we used the relation ~\cite{regge}
$D=- E\sigma^2$) \bee\label{61} \kern-50pt
S=\left(\begin{array}{cc}
-\frac G{E}    &\frac 1{\sigma E}\vspace{7mm}\\
-\frac 1{\sigma E}                     &\frac C{\sigma^2E}
\end{array}\right)=
\rme^{-\rmi\delta_2} \left(\begin{array}{cc}
\exp (\rmi\delta_1)\cos\alpha    &\sin\alpha\vspace{7mm}\\
-\sin\alpha                     &\exp (-\rmi\delta_1)\cos\alpha
\end{array}\right),
\eend where \bee\label{662} \delta_1=\arg f(\la,k), \hspace{10mm}
\delta_2=\arg f(\la,-k), \nonumber\\
\tan\alpha=\frac{2\sqrt{-k{\rm Im}\la}}{|f(\la,k)|}. \eend The two
scattering phases $\delta_1,\delta_2$ and the channel-mixing
nonelasticity angle $\alpha$ are real in sector IV, Im$\la<0,k>0$.
        The above definition of the $S-$matrix is subject to an arbitrariness.
         Without affecting
the unitarity and the meaning of the $S-$matrix elements we can
change the normalization by multiplying $\sigma$ by a unit length
complex number $\exp (\rmi \delta_3)$. Then, the off-diagonal
elements $S_{12}$ and $S_{21}$ in (\ref{61}) are multiplied by
$\exp (\mp\rmi\delta_3)$, resp. with the phase $\delta_3$
arbitrary. The values the angles (\ref{62}) take when $V=0$ may be
found in~\cite{shabad}.

  In sector IV the $S$-matrix elements $S_{11}$ and $S_{22}$ are no longer unit
length complex numbers, as they are in the cases of elastic
scattering in the
 outer world when $\la^2>0$, $k^2>0$ ($S_{22}=S_{12}=0$, $|S_{11}|=1$) or elastic scattering in the inner
world of Sec.~\ref{conf.} ($S_{11}=S_{12}=0$, $|S_{22}|=1$). Here
the unitarity can only be formulated with the inclusion of the
elements $S_{12}, S_{21}$ responsible for transitions between the
two channels as it was
 done above.
It would  be inappropriate  to be taking into account the
nonelasticity of the scattering process in sector IV by analytic
continuation with respect to $k$ or $\la$ from any of the elastic
sectors, as it is prescribed within the approach of
Ref.~\cite{alliluev}. The analytic continuation makes the
corresponding scattering phase complex but is unable to create the
lacking phase and the nonelasticity angle: a description of the
system with a greater number of degrees of freedom cannot be
achieved by mere analytic continuation. We have seen in this
Section that the singular \Sch is no longer a single-particle
equation.

Note, that in the present subsection we did not refer to any definite
 choice of boundary conditions. The only nesessary condition for the 
 $S$-matrix to be defined is that the fundamental solutions in (\ref{58})
 may be interpreted as asymptotically-free particles at the both end of 
 the interval $(0,\infty)$, which is guaranteed in our procedure with 
 any boundary conditions from the self-adjoint class. When defining (\ref{S})
 we acted differently from what we did when handlind sector III. Namely, unlike
 (\ref{SIII}), we took in (\ref{58}) the whole $\phi_{\pm\la,k}(r)$, and
  not its $\delta$-function-normalizable part. The two differ by the factor
   $R^{\pm\rmi\il}$. For a more general choice of boundary conditions, resulting
   from the substitution of arbitrary parameter $r_0$ for $R$ in (\ref{boundIV}),
   we would encounter the arbitrariness due to the factor $r_0^{\pm\rmi\il}$
   in the "noninvariant" $S$-matrix, defined like (\ref{SIII}). This is only 
   one-parameter dependence on the way of self-adjoint extension, whereas the 
   general case is $U$(2).
\subsection{Standard Liouville form}
Let us perform the following transformation of the wave
 function and of
the variable in the \Sch ($cf.$~\cite{Kamke}) in sector IV
 \bee\label{35}
 \widetilde{\psi}(\xi)=\psi(r)\left(\frac{1}{r^2}+\frac 1{R^2}\right)^\frac 1{4},
 \eend
 \bee\label{36}
 \xi(r)=\int_{|\frac\il{k}|}^{~r}\left(\frac{1}{r^2}+\frac 1{R^2}\right)^\frac 1{2}\rmd r=
 \int_1^{~y}\frac{\rmd y}{y}\left(1+y^2\right)^\frac 1{2},\nonumber\\
 y=r\left(\frac{-k}{\il}\right)=\frac r{R}>0.
 \eend
Note, that unlike \cite{Kamke}, \cite{naimark} the lower limit of
integration in the transformation (\ref{36})  is not the lower end
of the interval, on which the differential equation is defined.
 This transformation cannot be used in sector
  III, because the expression under the root signs in (\ref{35}), (\ref{36})
   may vanish for $r\in (0,\infty )$ there.

   The probability flux (\ref{27a}) is form-invariant under
   the transformations (\ref{35}), (\ref{36}).

   The \Sch (\ref{4}), (\ref{5}) takes the standard Liouville form
   \bee\label{37}
 -\frac{\rmd^2\widetilde{\psi}(\xi)}{\rmd\xi^2}+U(R;\xi)
 \widetilde{\psi}(\xi)= (\il )^2
\widetilde{\psi}(\xi).
  \eend
The potential in (\ref{37}) is
 \bee\label{38}
 U(R;\xi)=U_0(\xi)+U_V(R;\xi),\nonumber\\
  U_0(\xi)=\frac {1+6y^2}{4\left(1+y^2\right)^3}
  -\frac 1{4(1+y^2)},  \qquad \nonumber\\
  U_V(R;\xi)=
  \frac{y^2}{1+y^2}~R^2~V
  (y R).
  \eend
  Here $y$ is a function of  $\xi$
   to be obtained by inverting (\ref{36}). This function $y(\xi)$
   does not depend on the parameters $k,~\la$. Consequently, when
   $V=0$, the potential $U(R,\xi)=U_0(\xi)$ does not depend on them
   either. If, however, $V\neq 0$, the potential $U(R,\xi)$ depends
   upon the combination of the parameters $|k/\il|$, denoted as
   $|R|$ in the previous subsections.

\begin{figure}[htb]
  \begin{center}
       \includegraphics[bb = 0 0 405 210,
    scale=1]{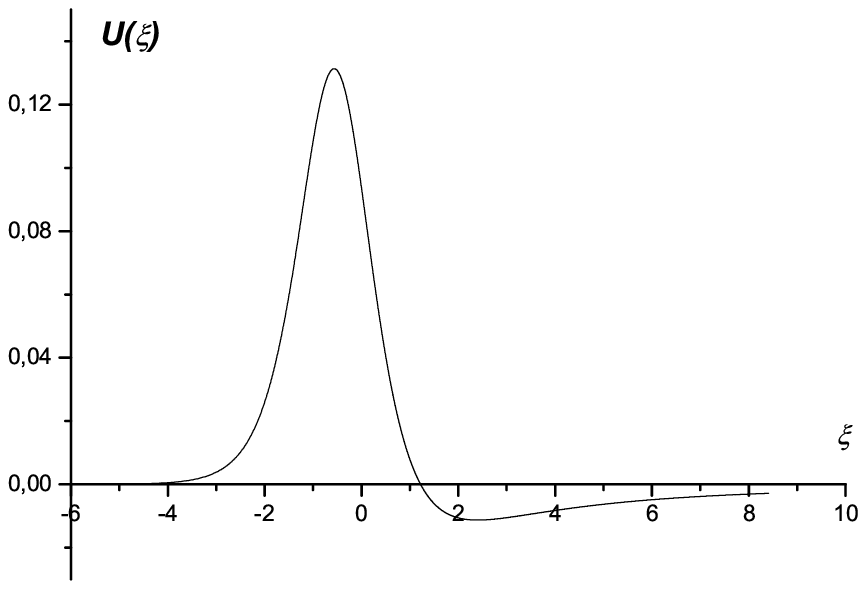}
    \caption{The effective barrier potential for confinement/
    deconfinement processes in the case of $V=0$. $~ U(\infty)=
     U(-\infty)=0$. The maximum is achieved at the value of $r$,
determined by the dimensional parameter $|R|$: $r_{max}=
     (5-21^{1/2})|R| ~~(\xi_{max}=-0.56)$}
     \label{fig:2}
     \end{center}
     \end{figure}

For the case $V(r)=0$ the effective barrier potential $U(\xi)$ is
plotted in Fig.~\ref{fig:2}. The inclusion of the potential $V\neq
0$, subject to conditions (\ref{poten}), does not affect the
asymptotic values $U(\pm\infty)$.
Near $r=\infty$ the variable $\xi$ becomes $\xi=r/|R|$,
$U(\infty)=0$, and equation (\ref{37}) takes the usual asymptotic
form of the \Sch \bee\label{39}
-\frac{\rmd^2\widetilde{\psi}}{\rmd r^2}=k^2\widetilde{\psi},
\quad \widetilde{\psi}=\sqrt{\frac{|k|}{\il}}~ \psi, \eend
describing particles, free in the infinitely remote region. Near
$r=0$ the barrier potential is $U(-\infty)=0$, the variable $\xi$
is $\xi=\ln(r/|R|)$, and equation ~(\ref{37}) is an equation for
particles, free in the region remote to the negative infinity,
with the solution given as (\ref{1.3}). \bee\label{free}
-\frac{\rmd^2\widetilde{\psi}(\xi)}{\rmd
\xi^2}=(\il)^2\widetilde{\psi}(\xi), \quad
\widetilde{\psi}(\xi)=\frac \psi {\sqrt r}.\eend Thus, we face
one-dimensional \Sch within infinite limits with the
 barrier potential that decreases at the both sides. It introduces the pattern
  of reflection and transmission, which we have studied above in this section
  directly in the primary variable $r$, without appealing to the transformation
(\ref{35}), (\ref{36}).

For large $r_U$ and small $r_L$   the transformation (\ref{36})
maps the box (\ref{box}) $r\in (r_L, r_U)$  into the box in the
$\xi$-space:
\begin{equation}\label{boxiv}
-\xi_L\leq \xi\leq \xi_U,\qquad \xi_L\gg1,\quad \xi_U\gg1 ,
\end{equation}
where the limits $\xi_L=-\xi(r_L)=-\ln (r_L/|R|),
~\xi_U=\xi(r_U)=r_U/|R|$ are meant to be independent of the
parameters $\la$ and $k$, and are connected with the limits
$r_L,~r_U$ in the initial space according to (\ref{limits}). In
the same asymptotic regime the boundary conditions (\ref{boundIV})
are transformed into the following periodic (antiperiodic)
boundary conditions in the box (\ref{boxiv}): \bee\label{boundxi}
\widetilde{\psi}(-\xi_L)=\pm\widetilde{\psi}(\xi_U),
 \nonumber\\
\left.\frac{\rmd\widetilde{\psi}(\xi)}{\rmd\xi}\right|_{\xi=-\xi_L}=\pm
\left.\frac{\rmd\widetilde{\psi}(\xi)}{\rmd\xi}\right|_{\xi=\xi_U}.
\eend The self-adjointness of the eigenvalue problem (\ref{37}) in
the box (\ref{boxiv}) with the periodic (antiperiodic) boundary
conditions (\ref{boundxi}), with the potential (\ref{38}), which
may at the most depend on the ratio $\la/k$, and with the ends of
the interval independent of $\la$ and $k$, is evident.

The eigenfunctions of the boundary problem (\ref{37}),
(\ref{boundxi}) are expressed in terms of the eigenfunctions
(\ref{eigenfunc}), (\ref{eigenfunc1}) as\bee\label{eigentilde}
\widetilde{\psi}^{(j)}_{\la,R}(\xi)=\left(\frac 1{R^2}+\frac
1{r^2}\right)^{\frac 1{4}}\psi^{(j)}_{\la,R}(r),\eend the spectrum
being given by solutions of  equations (\ref{prefinal}). The
functions are real. The set (\ref{eigentilde}) is complete. The
scalar product in the corresponding Hilbert space is defined with
the plane measure d$\xi$:\bee\label{producttilde}
(\widetilde{\psi}_1(\xi),\widetilde{\psi}_2
(\xi))\equiv\int_{-\xi_L}^{\xi_U} (\widetilde{\psi}_1(\xi))^*
\widetilde{\psi}_2(\xi){\rmd}\xi=\nonumber\\
=\int_{r_L}^{r_U}(\psi_1(r))^*\psi_2(r)\rmd\mu(r)\equiv(\psi_1(r),\psi_2(r)).\eend
The intermediate equality in (\ref{producttilde}) is proved using
(\ref{35}), (\ref{36}), (\ref{measureIV}). An arbitrary
sufficiently smooth function $\widetilde {F}(\xi)$ of $\xi$ from
$L^2(-\infty,\infty)$, $i.~e.$ such that
$\int_{-\infty}^\infty|\widetilde{F}(\xi)|^2\rmd\xi<\infty$ is
expanded into the generalized Fourier series over eigenfunctions
(\ref{eigentilde}) as\bee\label{Fouriertilde}
\widetilde{F}(\xi)=\frac{2\pi\sqrt R}{\xi_L+\xi_U}
\sum_{j=1,2}\sum_{~n=1}^{\infty}
C(\la_n^{(j)},R)\widetilde{\psi}_{\la_n,R}^{(j)}(\xi).\eend Owing
to the orthogonality relations (\ref{ortnorm}), which are the same
for the eigenfunctions $\widetilde{\psi}^{(j)}_{\la_n,R}(\xi)$ due
to (\ref{producttilde}), the expansion coefficients above are
\bee\label{coeftilde} C(\la_n^{(j)},R)=\frac 1{\sqrt R}
\int_{-\xi_L}^{\xi_U}\widetilde
{F}(\xi)(\widetilde{\psi}_{\la_n,R}^{(j)})^*\rmd\xi.\eend This is
equal to (\ref{coefficients}), provided that one
identifies\bee\label{Ftrans} \widetilde {F}(\xi)=F(r)\left(
1+\frac{R^2}{r^2}\right)^{\frac 1{4}}.\eend Both functions
$\widetilde {F}(\xi)$ and $F(r)$ are arbitrary and cannot depend
upon the parameter $R$, which only characterizes the set of
eigenfunctions used in the expansion. This statement does not
contradict to (\ref{Ftrans}),~ since the factor $(1+R^2/{r^2})$
does not contain $R$ after it is transformed to the variable $\xi$
according to (\ref{36}).

 The eigenfunction (\ref{eigentilde}) behaves near
$\xi=-\infty$ as\bee\label{ximin}
\widetilde{\psi}^{(j)}_{\la,R}(\xi)\asymp \frac 1{\sqrt\pi}\sin
\left(\frac{\mp\varepsilon^{(j)}\alpha+\delta_1 +\delta_2}{2}-\ln
R~\il-\xi\il\right), \eend and near $\xi=\infty$
as\bee\label{xiinfty} \widetilde{\psi}^{(j)}_{\la,R}(\xi)\asymp
\frac{\pm\varepsilon^{(j)}}{\sqrt\pi}\sin
\left(\frac{\pm\varepsilon^{(j)}\alpha+\delta_1
-\delta_2}{2}-\xi\il\right).\eend The logarithm of the dimensional
parameter $R$ is cancelled by an analogous logarithm that is
contained in the sum $\delta_1+\delta_2$, involved in
(\ref{ximin}), while the difference $\delta_1-\delta_2$,~ involved
in (\ref{xiinfty}),~ does not contain such logarithm.

For convergence of (\ref{Fouriertilde}) it is needed that
$\widetilde{F}(\pm\infty)=0$. In view of (\ref{Ftrans}) this
requirement is in agreement with (\ref{converge}).

\noindent \textbf{Acknowledgments.} The author thanks I.V.Andreev, I.B.Khriplovich,
S.L.Lebedev, I.V.Tyutin, and B.L.Voronov
for discussions. The work
was supported  in part by the Russian
 Foundation for Basic Research
 (project no 02-02-16944) and the President of the Russian Federation
 Program for Support of Leading Scientific Schools (grant no.LSS-1578.2003).

\end{document}